\documentclass[prb,reprint,noeprint,floatfix,citeautoscript]{revtex4-1}

\usepackage{cancel}
\usepackage{amsmath}
\usepackage{physics}
\usepackage{graphicx}
\usepackage{amssymb}
\usepackage{amsthm}
\usepackage{bm}
\usepackage{dcolumn}
\usepackage{braket}
\usepackage{longtable}
\usepackage{ragged2e}
\usepackage{txfonts}

\usepackage{color}
\usepackage[usenames,dvipsnames]{xcolor}
\definecolor{myblue}{rgb}{0,0,1}
\usepackage[breaklinks=true,colorlinks=true,linkcolor=myblue,urlcolor=myblue,citecolor=myblue]{hyperref}

\let\vr\undefined
\newcommand{\vr}{{\bm{r}}}

\begin{document}

\title{Vertex corrections to the polarizability do not improve
the $GW$ approximation for the ionization potential of molecules}

\author{Alan M.~Lewis}
\affiliation{Department of Chemistry and James Franck Institute,
University of Chicago, Chicago, Illinois 60637, USA}
\author{Timothy C.~Berkelbach}
\affiliation{Department of Chemistry, Columbia University, New York, New York 10027 USA}
\affiliation{Center for Computational Quantum Physics, Flatiron Institute, New York, New York 10010 USA}
\email{tim.berkelbach@gmail.com}


\begin{abstract}

The $GW$ approximation is based on the neglect of vertex corrections, which
appear in the exact self-energy and the exact polarizability. Here, we
investigate the importance of vertex corrections in the polarizability only.  We
calculate the polarizability with equation-of-motion coupled-cluster theory with
single and double excitations (EOM-CCSD), which rigorously includes a large
class of diagrammatically-defined vertex corrections beyond the random phase
approximation (RPA).  As is well-known, the frequency-dependent polarizability
predicted by EOM-CCSD is quite different and generally more accurate than that
predicted by the RPA.  We evaluate the effect of these vertex corrections on a
test set of 20 atoms and molecules. When using a Hartree-Fock reference,
ionization potentials predicted by the $GW$ approximation with the RPA
polarizability are typically overestimated with a mean absolute error of 0.3~eV.
However, those predicted with a vertex-corrected polarizability are typically
underestimated with an increased mean absolute error of 0.5~eV.  This result
suggests that vertex corrections in the self-energy cannot be neglected, at
least for molecules.  We also assess the behavior of eigenvalue self-consistency
in vertex-corrected $GW$ calculations, finding a further worsening of the
predicted ionization potentials.

\end{abstract}

\maketitle

\section{Introduction}

The $GW$ approximation~\cite{Hedin1965} has been widely and successfully used to
calculate the charged excitation energies associated with electron addition and
removal.  It has been applied to a variety of solids, including simple metals,
semiconductors, and transition metal
oxides,~\cite{Hedin1965,Strinati1980,Strinati1982,Hybertsen1985,Hybertsen1986,Godby1988,Aryasetiawan1998,Aulbur2000,Huser2013,Ergonenc2018}
and more recently to atoms and
molecules~\cite{Rostgaard2010,Bruneval2012,vanSetten2013,Bruneval2013,Huser2013,VanSetten2015,Caruso2016,Hung2017,Maggio2017,Govoni2018}.
Since its introduction, a number of attempts have been made to improve upon the
$GW$ approximation through the inclusion of diagrammatically-defined vertex
corrections beyond the random phase approximation (RPA).  In some cases, vertex
corrections are found to improve the accuracy of predicted excitation
energies~\cite{Shishkin2007,Gruneis2014,Chen2015,Schmidt2017}.  However, in
other cases, the lowest-order vertex corrections produce results that are only
marginally different, in both the condensed
phase~\cite{Rice1965,Minnhagen1974,Mahan1989,DelSole1994,deGroot1996,Shirley1996}
and in isolated molecules~\cite{Hung2017,Maggio2017,Ma2019}.

Here, we implement a large class of infinite-order vertex corrections to the
polarizability using equation-of-motion coupled-cluster theory with single and
double excitations (EOM-CCSD).  In addition to the particle-hole ring diagrams
resummed by the RPA, EOM-CCSD includes particle-hole ladder diagrams, 
particle-particle ladder diagrams, exchange diagrams, and
mixtures of all of the above. Furthermore, similar to the conventional $GW$-based implementation
of the Bethe-Salpeter equation~\cite{Rohlfing2000,Onida2002}, the propagator
lines are dressed and particle-hole interactions are screened.  We use this
improved polarizability to construct a more accurate screened Coulomb
interaction $W$, for use in the $GW$ approximation; because this style of vertex
corrections aims to calculate $W$ in terms of the response of a test charge due
to a test charge, it is sometimes referred to as $G_0W^\mathrm{tc-tc}$.  We assess 
this vertex-corrected $GW$ approximation by calculating the ionization
potentials of the twenty smallest atoms and molecules of the $GW100$ test set,
which has recently been introduced for the purpose of benchmarking different
implementations of the $GW$
approximation.\cite{VanSetten2015,Caruso2016,Maggio2017,Govoni2018} By comparing
our results to those obtained using conventional RPA screening, we conclude that
vertex corrections to the polarizability worsen the accuracy of the $GW$
approximation for ionization potentials of molecules.  We also implement eigenvalue self-consistency in
our vertex-corrected $GW$ calculations, and again find no improvement.  We
conclude that high-order vertex corrections to the structure of the self-energy are required to
improve on existing methods.

\section{Theory}

Charged excitation energies, associated with electron addition and removal, can be
calculated by finding the poles of the one-particle Green's function,
\begin{equation}
G(1,2) = - i\mel{\Psi_0}{T[\psi^{\dagger}(1)\psi(2)]}{\Psi_0}.
\label{GF}
\end{equation}
Here $\psi^{\dagger}$ and $\psi$ are field operators, the labels 1 and 2
indicate a set of position and time variables, i.e.~$1=(\vr_1,t_1)$, $T$ is the time-ordering
operator, and $\ket{\Psi_0}$ is the ground state of the many-electron
system. In practice, $G$ is usually calculated via the self energy $\Sigma$,
defined by the Dyson equation,
\begin{equation}
G(1,2) = G_0(1,2) + \int G_0(4,2) \Sigma(3,4) G(1,3) d(3) d(4),
\label{Hedin1}
\end{equation}
where $G_0$ is a noninteracting or mean-field Green's function.
If $G_0$ is chosen to be the Hartree Green's function, then the exact
self-energy may be written as~\cite{Hedin1965}
\begin{equation}
\Sigma(1,2) = i \int G(1,4) W(1,3) \Gamma(4,2,3) d(3) d(4),
\label{HedinSigma}
\end{equation}
where $W$ is the screened Coulomb interaction and $\Gamma$ is a three-point
vertex function. The screened Coulomb interaction 
is given by
\begin{equation}
\begin{split}
W(1,2) &= v(1,2) + \int v(4,2) \Pi^\star(3,4) W(1,3) d(3) d(4) \\
    &= v(1,2) + \int v(4,2) \Pi(3,4) v(1,3) d(3) d(4),
\label{HedinW}
\end{split}
\end{equation}
where $v(1,2) = |\vr_1-\vr_2|^{-1}\delta(t_1-t_2)$ is the usual Coulomb
interaction.
In the screened Coulomb interaction, $\Pi^\star$ and $\Pi$ are the irreducible and
reducible polarizabilities,
\begin{align}
\label{eq:PiStar}
\Pi^\star(1,2) &= -i \int G(2,3) G(4,2) \Gamma(3,4,1) d(3) d(4), \\
\label{eq:Pi}
\Pi(1,2) &= 
        -i\langle \Psi_0 | T [ \tilde{\rho}(1) \tilde{\rho}(2) |\Psi_0\rangle,
\end{align}
where $\tilde{\rho} = \rho - \langle \Psi_0 | \rho |\Psi_0\rangle$ and
$\rho = \psi^\dagger\psi$.
The three-point vertex function $\Gamma$ appearing in Eqs.~\eqref{HedinSigma}
and \eqref{eq:PiStar} is defined by
\begin{equation}
\begin{split}
\Gamma(1,2,3) &= \delta(1,2)\delta(1,3) \\
    &\hspace{1em} + \int 
        \frac{\delta \Sigma(1,2)} {\delta G(4,5)} G(4,6) G(7,5) \\
    &\hspace{3em} \times \Gamma(6,7,3) d(4) d(5) d(6) d(7).
\end{split}
\label{HedinGamma}
\end{equation}
The conventional $GW$ approximation follows by setting
$\Gamma(1,2,3) = \delta(1,2)\delta(1,3)$, i.e.~neglecting
vertex corrections, leading to
\begin{align}
\Sigma(1,2) &\approx i G(1,2) W(1,2), \\
\Pi^\star(1,2) &\approx -i G(2,1) G(1,2).
\label{eq:PiRPA}
\end{align}
In practice, the $GW$ approximation is commonly implemented without
self-consistency, where $G$ and $W$ are evaluated in a one-shot manner
based on the mean-field starting point, leading to the so-called
$G_0 W_0$ approximation.   

\begin{figure}[t]
\centering
\includegraphics[scale=1.0]{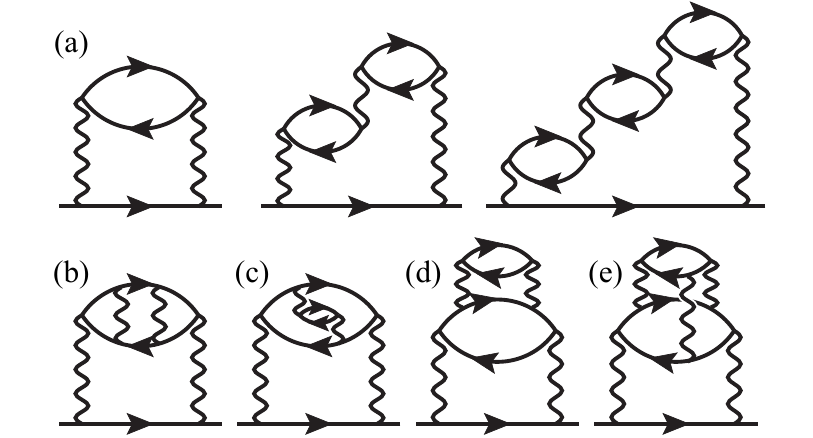}
\caption{Example self-energy diagrams included when the polarizability is
calculated with EOM-CCSD.  The diagrams in (a) are those included in the usual
$GW$ approximation with RPA screening.  Diagram (b) is included with a TDHF
polarizability, diagrams (c) and (d) would be included with a $GW$-based BSE
polarizability, and diagram (e) -- showing an example hole-hole ladder
interaction -- is only included with the EOM-CCSD polarizability.
}
\label{fig:diagrams}
\end{figure}

The exact \textit{reducible} polarizability given in Eq.~\eqref{eq:Pi} has a Lehmann
representation 
\begin{equation}
\begin{split}
\Pi(1,2) &= -i\theta(t_1-t_2) \sum_{n>0} e^{-i\Omega_n(t_1-t_2)} 
                \rho_n(\vr_1) \rho_n^*(\vr_2) \\
    &\hspace{1em} + (1\leftrightarrow 2)
\end{split}
\end{equation}
where $\rho_n(\vr) = \langle \Psi_0 | \rho(\vr) |\Psi_n\rangle$ 
and $\Omega_n = E_n-E_0$.
We note that the reducible polarizability is closely related to a certain time-ordering of the
two-particle Green's function~\cite{FetterWalecka}.
Separating the $GW$ self-energy into its exchange and correlation
components gives
\begin{subequations}
\begin{align}
\Sigma^\mathrm{x}(1,2) &= i G(1,2) v(1,2) \\
\Sigma^\mathrm{c}(1,2) &= i G(1,2) \int v(4,2) \Pi(3,4) v(1,3) d(3) d(4).
\end{align}
\end{subequations}
In a finite single-particle basis, the frequency dependence can be treated 
analytically such that the correlation component of the self-energy is given by
\begin{equation}
\begin{split}
\Sigma^\mathrm{c}_{pq}(\omega) &= 
    \sum_{n>0} \Bigg[ \sum_i \frac{(pi|\rho_n^*)(\rho_n|iq)}
                {\omega - (\varepsilon_i - \Omega_n) -i\eta} \\
        &\hspace{3em} + \sum_a \frac{(pa|\rho_n)(\rho_n^*|aq)}
            {\omega-(\varepsilon_a+\Omega_n) + i\eta} \Bigg]
\label{eq:SigmaLehmann}
\end{split}
\end{equation}
where
\begin{equation}
(pq|\rho_n) = \int d\vr_1 \int d\vr_2 \phi_p^*(\vr_1) \phi_q(\vr_1)
    |\vr_1-\vr_2|^{-1} \rho_n(\vr_2).
\end{equation}
Here and throughout we use indices $i,j$ to denote orbitals that are
occupied and $a,b$ to denote orbitals that are unoccupied in the mean-field
reference determinant.
Equation~\eqref{eq:SigmaLehmann} provides the formalism by which any
theory of the polarizability can be employed in the $GW$ approximation. 
For example, conventional RPA screening (no vertex corrections),
as defined by Eq.~\eqref{eq:PiRPA},
is recovered if 
the excitation energies
$\Omega_n$ are obtained from the familiar eigenvalue
problem~\cite{RingSchuck}
\begin{equation}
\label{eq:rpa}
\left(
\begin{array}{cc}
\mathbf{A}   & \mathbf{B} \\
-\mathbf{B}^* & -\mathbf{A}^*
\end{array}
\right)
\left(
\begin{array}{c}
\mathbf{X} \\
\mathbf{Y}
\end{array}
\right)
= 
\left(
\begin{array}{c}
\mathbf{X} \\
\mathbf{Y}
\end{array}
\right) \mathbf{\Omega},
\end{equation}
where
\begin{subequations}
\label{eq:rpa_ints}
\begin{align}
A_{ia,jb} &= (\varepsilon_a - \varepsilon_i) \delta_{ab}\delta_{ij} + \langle ib | aj \rangle, \\ 
B_{ia,jb} &= \langle ij | ab \rangle,
\end{align}
\end{subequations}
two-electron integrals are defined by
\begin{equation}
\langle pq|rs\rangle = \int d\vr_1\int d\vr_2 
    \phi_p^*(\vr_1) \phi_q^*(\vr_2) |\vr_1-\vr_2|^{-1} \phi_r(\vr_1) \phi_s(\vr_2),
\end{equation}
and
the transition moments $\rho_n(\vr)$ are given by 
\begin{equation}
\rho_n(\vr) = \sum_{ai} 
    \left[ X_{ia}^{(n)} \phi_a(\vr)\phi_i^*(\vr)
        + Y_{ia}^{(n)} \phi_i(\vr)\phi_a^*(\vr) \right]
\end{equation}
with the orthonormalization condition
\begin{equation}
\sum_{ai} \left\{ [X_{ai}^{(m)}]^*X_{ai}^{(n)} 
        - [Y_{ai}^{(m)}]^*Y_{ai}^{(n)} \right\}
    = \delta_{nm}.
\end{equation}
This flavor of RPA is sometimes referred to as ``direct RPA'' because it
neglects the exchange integrals that would arise from antisymmetrization in
Eqs.~\eqref{eq:rpa_ints}.  If the antisymmetrized integrals $\langle
pq||rs\rangle \equiv \langle pq|rs\rangle - \langle pq|sr\rangle$ are
maintained, then the screening is equivalent to time-dependent Hartree-Fock
(TDHF). This level of theory was used to implement vertex corrections in the
recent work of Maggio and Kresse~\cite{Maggio2017}, which will also be tested
here.

\begin{figure*}[t]
\centering
\includegraphics[scale=1.0]{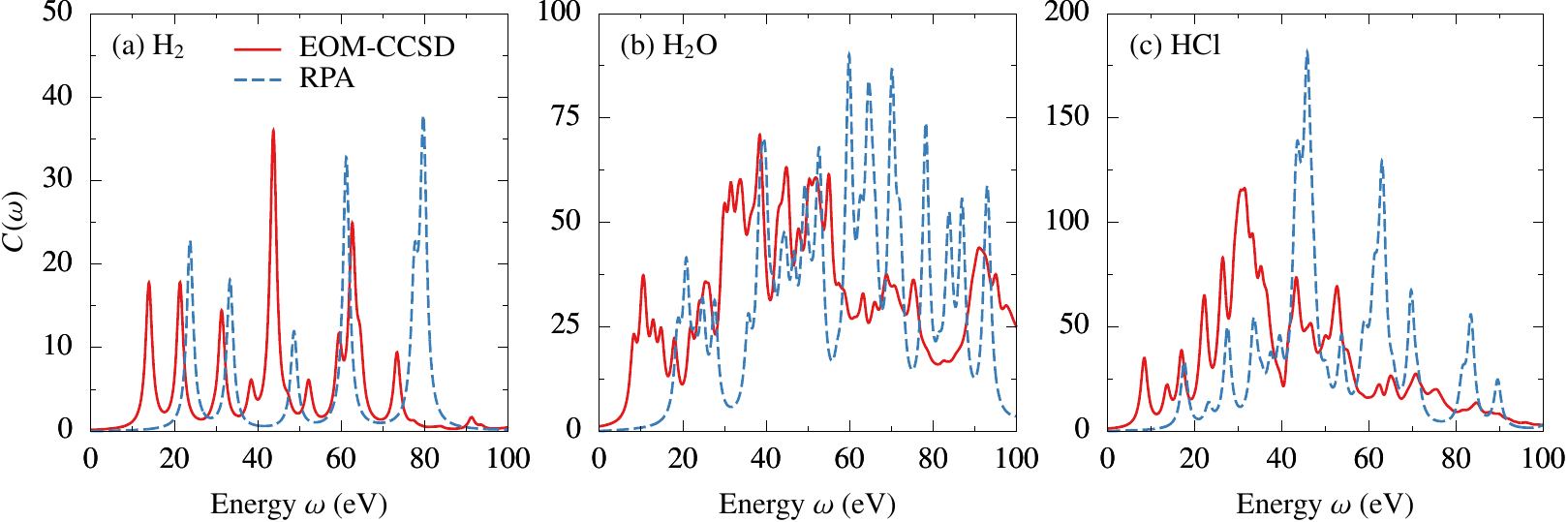}
\caption{The spectral function of the polarizability $C(\omega)$ for H$_2$,
H$_2$O, and HCl calculated using EOM-CCSD and the RPA. All calculations
are done in the def2SVP basis using a Hartree-Fock reference and 
using a numerical broadening of 1~eV.}
\label{fig:Density}
\end{figure*}

Here, we implement vertex corrections in the polarizability by using EOM-CCSD
transition densities and excitation energies in Eq.~\eqref{eq:SigmaLehmann}.
The formalism rigorously subsumes RPA screening (no vertex corrections) and
TDHF screening.
Briefly, EOM-CCSD excitation energies are defined as eigenvalues of the
similarity-transformed $\bar{H} = e^{-T} H e^{T}-E_{\mathrm{CCSD}}$ in the
subspace of determinants that are singly and doubly excited with respect to a
reference determinant $|\Phi\rangle$.  The operator $T$ creates single and
double excitations, $T = \sum_{ai} t_i^a a_a^\dagger a_i + \frac{1}{4}
\sum_{abij} t_{ij}^{ab} a_a^\dagger a_b^\dagger a_j a_i$, and the amplitudes are
determined by the nonlinear system of equations $\langle \Phi_i^a |e^{-T}He^{T}
|\Phi \rangle = 0$ and $\langle \Phi_{ij}^{ab} |e^{-T}He^{T} |\Phi \rangle = 0$.
The right-hand eigenstates of $H$ are then given by
\begin{subequations}
\begin{align}
|\Psi_0\rangle &= e^{T} |\Phi\rangle \\
|\Psi_n\rangle &= \left[ r_0 + \sum_{ai} r_i^a a_a^\dagger a_i 
    + \frac{1}{4} \sum_{abij} r_{ij}^{ab} a_a^\dagger a_b^\dagger a_j a_i 
        \right] e^{T} |\Phi\rangle
\end{align}
\end{subequations}
and the left-hand eigenstates ($n \ge 0$) by
\begin{equation}
\langle\tilde{\Psi}_n| = \langle \Phi| \left[l_0 + 
    \sum_{ai} l_a^i a_i^\dagger a_a
    + \frac{1}{4} \sum_{abij} l_{ab}^{ij} a_i^\dagger a_j^\dagger a_b a_a
    \right] e^{-T}.
\end{equation}
The transition densities follow naturally
\begin{subequations}
\begin{align}
\rho_n(\vr) &= \sum_{pq} \phi_p^*(\vr) \phi_q(\vr)
    \langle \tilde{\Psi}_0 | a_p^\dagger a_q |\Psi_n\rangle \\
\rho_n^*(\vr) &= \sum_{pq} \phi_p^*(\vr) \phi_q(\vr)
    \langle \tilde{\Psi}_n | a_p^\dagger a_q | \Psi_0\rangle,
\end{align}
\end{subequations}
for which analytic expressions can be simply obtained~\cite{Stanton1993}.

EOM-CCSD is universally viewed as superior to the HF-based RPA for electronic
excitation energies of molecules.  For excited states that are well-described as
single excitations, EOM-CCSD is accurate to about
$0.1$--$0.3$~eV,~\cite{Schreiber2008} whereas the HF-based RPA displays errors 
of 1~eV or more~\cite{Wiberg2002}.
Improved results can be obtained with alternative choices of the 
mean-field reference,
inclusion or exclusion of exchange, or in combination with time-dependent
density functional theory~\cite{Wiberg2002,Sottile2005,Caricato2011,Ma2019}.  
In a more rigorous sense, the RPA can be derived as an approximation to
EOM-CCSD, as recently discussed by one of us~\cite{Berkelbach2018}.
Diagrammatically, the EOM-CCSD polarizability resums all particle-hole ring
diagrams (as in the RPA), as well as particle-particle, hole-hole, and
particle-hole ladder diagrams, exchange diagrams, and mixtures of all of the
above.  These extra diagrams define the class of vertex corrections included in
the polarizability beyond the RPA.  When the RPA or EOM-CCSD polarizability is
used in the non-self-consistent $GW$ approximation, we will term the method the
$G_0W_0$ or $G_0W_{\mathrm{CC}}$ approximation, respectively.  In
Fig.~\ref{fig:diagrams}, we show some example self-energy diagrams included with
an EOM-CCSD polarizability, and identify some that are included in various lower
levels of theory.

\section{Results}

In the results to follow, we study atoms and molecules from the
$GW$100 test set~\cite{VanSetten2015}.  Due to the relatively high computational
cost of obtaining many highly-excited states via EOM-CCSD, we
only consider the smallest twenty atoms and molecules, using the
polarized double-zeta def2SVP basis set~\cite{Weigend2005,Weigend2006}.
Although we have not optimized the performance, the calculation
of ionization potentials with EOM-CCSD vertex corrections
in the polarizability can be performed in a manner that scales as $N^7$.
By comparing results within a given basis set, our conclusions are largely
free of basis set incompleteness error but numerical values should not be
compared to experiment or to predictions in other basis sets.
To give a rough sense of basis set completeness, previous $G_0W_0$ calculations have shown that IPs
calculated in this basis set underestimate the complete basis set limit by about 
0.3--0.5~eV~\cite{vanSetten2013}.  We have performed the following calculations for the five
smallest atoms and molecules in the larger def2-TZVPP basis set, and find that our results and conclusions
are unchanged.  Where appropriate, we will also compare to
previously published results in larger basis sets, which demonstrate that our
general conclusions are robust.
All calculations were performed with the PySCF software
package~\cite{Sun2018}.

First, to illustrate the differences between the RPA and EOM-CCSD
polarizabilities, we consider the two-particle spectral function
\begin{equation}
C(\omega) = \sum_{n>0} \sum_{pq} 
    |\langle \Psi_n|a_p^\dagger a_q |\Psi_0\rangle|^2 \delta(\omega-\Omega_n),
\end{equation}
which is closely related to the imaginary part of the polarizability. This
two-particle spectral function contains
the same neutral-excitation quantities that enter into the $GW$ self-energy,
i.e.~the transition density matrix elements and the excitation energies.
In Fig.~\ref{fig:Density}, we show $C(\omega)$ for three example molecules, H$_2$
(for which EOM-CCSD is exact), H$_2$O and HCl, over a very wide spectral range.
RPA and EOM-CCSD calculations are done with a Hartree-Fock (HF) reference; see
below for further discussion of this choice.  Due to the very slow decay of
$(\omega-E)^{-1}$, the self-energy at a given frequency is affected by a very
large number of neutral excitation energies, as can be inferred from
Eq.~\eqref{eq:SigmaLehmann}. Indeed, truncating the number of neutral excitation
energies retained in the polarizability can affect the ionization potentials
(IPs) by anywhere from 0.1 to 1~eV~\cite{Bruneval2016}.  For all molecules, the
RPA spectra are shifted to higher energies by 10~eV or more.  This behavior is
because the RPA polarizability does not include the electron-hole ladder
diagrams that are included in the EOM-CCSD polarizability.  These ladder
diagrams reduce the excitation energy of molecules and lead to bound exciton
states in semiconductors~\cite{Hanke1975,Rohlfing2000,Onida2002}.  The
overestimation of excitation energies can be partially, but not systematically,
alleviated by choosing a mean-field reference with a smaller gap. Compared to
HF, essentially all flavors of density functional theory (DFT) satisfy this
property, which explains the popularity of the DFT+RPA approach.  Roughly
speaking, a larger spectral gap in the polarizability will reduce the screening,
such that the $GW$ correction to HF is less effective and the IPs are too large,
which is indeed observed in our $G_0W_0$@HF calculations.  Because the EOM-CCSD
polarizability has a smaller (more accurate) spectral gap, the screening is
stronger, the $GW$ correction is larger, and the IPs are significantly reduced
in magnitude.

\begin{table}[t]
\begin{tabular}{ c | c | c | c | c }
\hline\hline
Molecule & $\Delta$CCSD(T) & $G_0W_0$@HF & $G_0W_{\rm CC}$@HF & $G_{\rm ev}W_{\rm CC}$@HF \\
\hline
He & 24.31 & 24.32 & 23.82 & 23.78 \\
Ne & 21.08 & 20.98 & 20.32 & 20.19 \\
H$_2$ & 16.26 & 16.24 & 15.97 & 15.99 \\
Li$_2$ & 5.07 & 5.03 & 4.90 & 4.95\\
LiH & 7.69 & 7.81 & 6.95 & 6.54 \\
FH & 15.60 & 15.64 & 14.99 & 14.82 \\
Ar & 15.20 & 15.31 & 15.06 & 15.00 \\
H$_2$O & 12.07 & 12.27 & 11.66 & 11.53 \\
LiF & 10.76 & 10.51 & 9.22 & 8.52 \\
HCl & 12.15 & 12.31 & 12.05 & 12.05 \\
BeO & 9.98 & 9.63 & 8.65 & 8.11 \\
CO & 13.70 & 14.73 & 14.11 & 13.97 \\
N$_2$ & 15.27 & 16.98 & 16.69 & 16.68 \\
CH$_4$ & 14.25 & 14.51 & 14.10 & 14.03 \\
BH$_3$ & 13.17 & 13.42 & 13.03 & 12.98 \\
NH$_3$ & 10.32 & 10.61 & 10.10 & 10.00 \\
BF & 10.82 & 10.98 & 10.69 & 10.70\\
BN & 11.89 & 11.36 & 11.04 & 11.00 \\
SH$_2$ & 9.89 & 10.07 & 9.81 & 9.82 \\
F$_2$ & 15.56 & 16.03 & 15.30 & 15.02 \\
\hline
{ ME} & - & { +0.19} & { -0.31} & { -0.44} \\
{ MAE} & - & { 0.31} & { 0.52} & { 0.64} \\
\hline\hline
\end{tabular}
\caption{The first ionization potential in eV calculated using $G_0W_0$@HF,
$G_0W_{\rm CC}$@HF, and $G_\mathrm{ev}W_{\rm CC}$@HF, where EOM-CCSD is used to
calculate the screened interaction $W_{\rm CC}$. Errors are calculated with respect
to $\Delta$CCSD(T).
All calculations are done in the def2SVP basis using a Hartree-Fock reference.
}
\label{Data}
\end{table}

\begin{figure*}[t]
\centering
\includegraphics[scale=1.0]{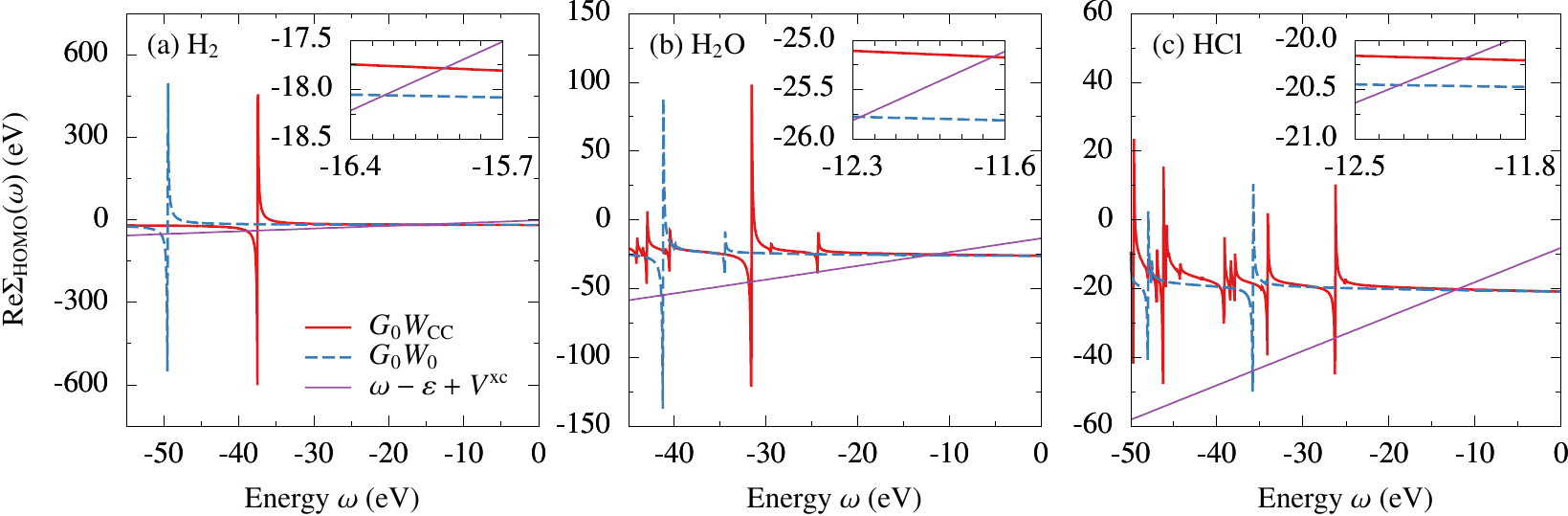}
\caption{The real part of the HOMO self-energy for H$_2$, H$_2$O, and HCl
calculated using the vertex-corrected $G_0W_{\textrm{CC}}$ and non-vertex-corrected 
$G_0W_0$ approximations. Each 
inset magnifies a (0.7~eV)$\times$(1~eV) region around the quasiparticle energies, where
$\Sigma_{\textrm{HOMO}}(\omega) = \omega - \varepsilon +V^{\mathrm{xc}}$. The
self-energy is calculated with a small imaginary part of $\eta = 0.03$~eV.}
\label{Sigma}
\end{figure*}

In Tab.~\ref{Data}, we present the first IP of the
twenty smallest atoms and molecules of the $GW$100 test set, obtained via
$G_0W_0$ and $G_0W_\mathrm{CC}$.  As a reference, we calculate
the first IP using $\Delta$CCSD(T), i.e.~as a 
difference in ground-state energies between the neutral and charged systems 
using CCSD with perturbative triple excitations.
In all $GW$ calculations, we use Hartree-Fock (HF) theory as the mean-field
reference, which has been established as a good choice for 
molecules~\cite{Caruso2016,Bruneval2013,Caruso2012}.
Importantly, the HF starting point has no self-interaction error through
first order.  However, the missing correlation and orbital relaxation
leads to HF IPs that are too large in magnitude (orbital energies
are too negative).
Consistent with previous results~\cite{Caruso2016}, the $G_0W_0$@HF
approximation predicts reasonably accurate IPs, with a mean error (ME) of
$+0.19$~eV and a mean absolute error (MAE) of $0.31$~eV (these
can be compared to identical calculations in the larger def2-TZVPP basis~\cite{Caruso2016},
which have a ME of $+0.26$~eV and a MAE of $0.35$~eV). The vertex-corrected
$G_0W_\mathrm{CC}$@HF approximation gives worse results and underestimates IPs, with a ME of
$-0.31$~eV and a MAE of $0.52$~eV.  In particular, the vertex-corrected
calculations give a less accurate IP for eleven of the twenty molecules.
We also note that for the two-electron molecules H$_2$ and He, the EOM-CCSD
polarizability, and thus $W$, is exact; however the results for both molecules
are worse when the exact $W$ is used in the $GW$ approximation.

This reduction in the IP can be understood from Fig.~\ref{Sigma}, which
shows the frequency dependence of the real part of the self-energy
for the highest occupied molecular orbital (HOMO), corresponding to the
first IP. 
The poles of the self-energy with vertex corrections are clearly shifted to
higher energies (less negative) by about $10$~eV, consistent with the
differences in the polarizabilities shown in Fig.~\ref{fig:Density}. The
pole strengths are relatively unchanged, and therefore the IPs are
reduced in magnitude, compared to those predicted by the $GW$ approximation
without vertex corrections.  

\begin{figure}[t]
\centering
\includegraphics[scale=1.0]{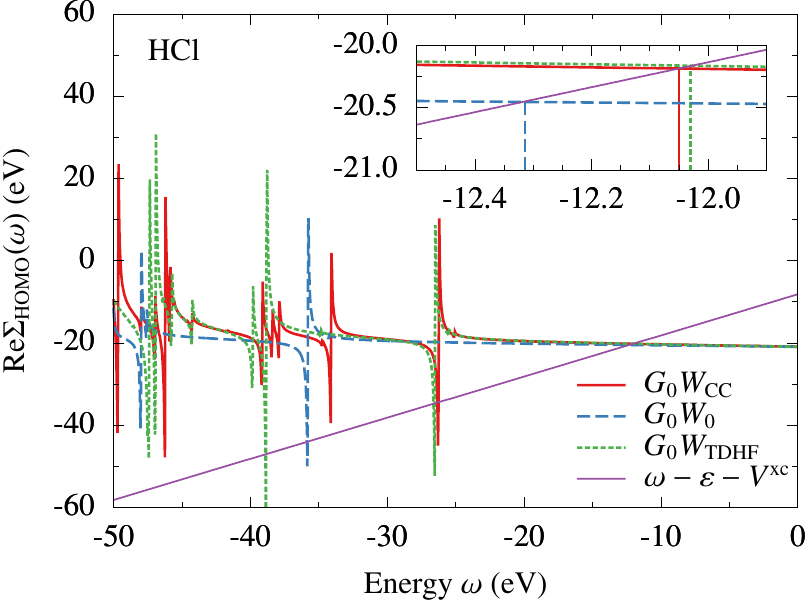}
\caption{
The same as in Fig.~\ref{Sigma}(c), but also including the result with
TDHF vertex corrections, corresponding to electron-hole interactions
in the polarizability.
}
\label{fig:tdhf}
\end{figure}

Although we do not show the detailed results here, we have also implemented
vertex corrections at the TDHF level~\cite{Maggio2017}, which can be viewed as
intermediate between the RPA (no vertex corrections) and EOM-CCSD.  TDHF vertex
corrections to the polarizability add particle-hole ladder diagrams -- shown in
Fig.~\ref{fig:diagrams}(b) -- that are responsible for excitonic effects and
expected to be important in molecules.  For the molecules considered here, the
TDHF excitation energies are quite close to those of EOM-CCSD, such that the IPs
predicted via the vertex-corrected $GW$ approximation are similar.  Specifically, the IPs
predicted with TDHF vertex corrections exhibit a mean error of $-0.28$~eV and a
mean absolute error of $0.50$~eV.  These can be compared to the analogous 
TDHF vertex-corrected results of Ref.~\onlinecite{Maggio2017} (there called $G_0W_0^{\mathrm{tc-tc}}$), 
which have a ME of $-0.06$~eV and a MAE of
$0.15$~eV.  Although these latter results appear more accurate than our own, the errors
are obtained by comparing $GW$ results extrapolated to the complete basis set limit to CCSD(T) results
in a finite cc-pVQZ basis set; as discussed by those authors~\cite{Maggio2017}, the CCSD(T) IPs
are likely underestimated by $0.10$--$0.15$~eV, such that a consistent comparison in the basis set limit
would worsen the performance of those vertex-corrected $GW$ calculations and bring them into better
agreement with our own (for which errors are consistently calculated in the same basis set).  

The $GW$ self energies calculated using the
RPA, TDHF, and EOM-CCSD polarizabilities for the HCl molecule are shown in
Fig.~\ref{fig:tdhf}.  The similarity between the TDHF and EOM-CCSD polarizabilities
can be understood based on the weakly correlated nature of the molecules
studied, as well as the dominant one-particle+one-hole nature of the low-energy 
excitations.  For molecules or solid-state materials with a small or vanishing gap,
the TDHF and EOM-CCSD polarizabilities are expected to differ more qualitatively
and may yield larger differences in ionization potentials when used with the vertex-corrected 
$GW$ approximation.

These collective results demonstrate that high-quality vertex corrections to the
polarizability do not improve the ionization potentials of small molecules
within the $GW$ approximation; when used with a HF reference, these vertex
corrections make the results worse by predicting IPs that are significantly too
small in magnitude.  However, we find that TDHF vertex corrections to the
polarizability, as recently implemented by Maggio and Kresse~\cite{Maggio2017}
for both the polarizability and the self-energy, are a good approximation to
those produced by the more expensive EOM-CCSD approach presented here and
represent a promising and affordable approach for weakly correlated, gapped materials.

These findings can be compared to
previous solid-state calculations, where it was found that adding low-order
vertex corrections to the polarizability alone unphysically reduced the
bandwidth\cite{Mahan1989} and increased the work function\cite{Morris2007} of
simple models of metals, and increased the quasiparticle energy of insulators
and semiconductors~\cite{DelSole1994}.  It has also been shown that small
improvements to the polarizability make little difference to the ionization
potentials of atoms~\cite{Morris2007}. The present work extends these previous
results by employing a far more accurate and diagrammatically-defined
polarizability, and demonstrating the behavior across a range of
molecular systems.

Having addressed the low-level RPA treatment of screening, we now mention the two
remaining sources of error in the $GW$ approximation: vertex corrections to the
self-energy and self-consistency.  The former are more challenging to implement
than vertex corrections to the polarizability, however future work will
address this issue.  While self-consistency is also challenging,
one relatively inexpensive option is to enforce eigenvalue
self-consistency~\cite{Hybertsen1986,Pavlyukh2004,Blase2011,Faber2011,Marom2012,Kaplan2015,Kaplan2016}. 
In this approach, the quasiparticle eigenvalues associated with each orbital are
replaced with the newly calculated quasiparticle energies after each iteration
of the $GW$ calculation until self-consistency is established. Despite not being
fully self-consistent, these methods have been found to significantly reduce the
starting point dependence of $GW$
calculations~\cite{Faber2011,Kaplan2015,Kaplan2016}.  Here, we implement and
test eigenvalue self-consistency for EOM-CCSD vertex-corrected $GW$
calculations.

A major advantage of using EOM-CCSD for vertex corrections is that the
coupled-cluster framework is extremely insensitive to the choice of
mean-field reference~\cite{ShavittBartlett}.  This can be understood by the Thouless
theorem, which shows that the single excitation part of the coupled-cluster
wave operator, $e^{T_1}$, is able to transform a Slater determinant into
any other~\cite{Thouless1960}.  This insensitivity is responsible for the common
choice of a HF reference, for which the working equations are simpler.  In
numerical tests, we find that eigenvalue self-consistency makes almost no change
to the EOM-CCSD polarizability, and thus we enforce eigenvalue self-consistency
in $G$ only (but the results should be understood as essentially those of
complete eigenvalue self-consistency).  We refer to this approach as 
$G_{\rm ev}W_{\rm CC}$; the IPs predicted by this method are listed in
Tab.~\ref{Data}.  We find that enforcing eigenvalue self-consistency further
deterioriates the accuracy, yielding a mean error of $-0.44$~eV and a mean
absolute error of $0.64$~eV.  We conclude that combining eigenvalue
self-consistency with a large class of vertex corrections to the polarizability
further worsens the $GW$ approximation, leading to IPs that are severely
underestimated in magnitude.

\section{Conclusion}

In this work, we have investigated the effect of high-quality vertex corrections
to the polarizability for use in the $GW$ approximation.
Vertex corrections were implemented using EOM-CCSD, which corresponds to
an infinite order resummation of particle-hole, particle-particle, and
hole-hole ladder diagrams, in addition to the usual ring diagrams, and mixtures
of all of the above~\cite{Berkelbach2018}.
The resulting polarizability is undeniably more accurate than that predicted
by the RPA. However, the vertex-corrected $GW$ approximation produces
\textit{worse} results than calculations without vertex corrections, when applied to a test set of 
twenty small atoms and molecules.  Specifically, the improved
treatment of screening correctly decreases the IPs, however it overcompensates and
predicts IPs that are significantly too small.
Enforcing eigenvalue self-consistency also showed no improvement.

We have focused on the use of the $GW$ approximation to predict the first IP,
even though the Green's function contains much more information.  It is possible that
the vertex corrections implemented in the polarizability would yield an improvement
in quantitites other than the principle IP.  For example, it can be clearly seen in
Fig.~\ref{fig:tdhf} that different treatments of screening leads to very different structure
in the self-energies at higher (more negative) energies, which will lead to significantly different
predictions of the locations of satellite peaks in the one-particle spectral function.
For example in HCl, although the TDHF and EOM-CCSD vertex corrections predict very similar 
quasiparticle and first satellite peaks, they predict a second satellite peak that differs by
about 5~eV.
However, in all of the molecules we checked, the weight of these satellite peaks is so small
as to be physically inconsequential.  It will be interesting to investigate the role of
vertex corrections on the satellite structure of molecules or materials with stronger electron correlation.

As mentioned above, the only remaining approximation
is the neglect of vertex corrections in the self-energy.  Without these, the
$GW$ approximation neglects the transient interactions between the screened
particle and the particle-hole pairs responsible for screening. Additionally,
the neglected exchange diagrams in the self-energy are responsible for a
self-screening error~\cite{Romaniello2009,Wetherell2018}.  However, when the
lowest-order vertex corrections to the self-energy were included in the
calculation of the bandgaps of silicon\cite{Bobbert1994} and a semiconducting
wire,\cite{deGroot1996} only small improvements were observed. Furthermore,
these corrections are found to cancel with the lowest order corrections to the
polarizability, as mentioned
previously.\cite{Rice1965,Minnhagen1974,Mahan1989,DelSole1994,deGroot1996,Shirley1996}
In order to systematically improve upon the $G_0W_0$ approximation, it appears
necessary to include high-order vertex corrections to both the self-energy and
the polarizability.

We note that a number of other Green's function based approaches include
infinite-order vertex corrections in both the self-energy and the
polarizability, including the two-particle-hole Tamm-Dancoff
approximation~\cite{Walter1981}, the third-order algebraic diagrammatic
construction (ADC(3))~\cite{Schirmer1983}, and the EOM-CC Green's
function~\cite{Nooijen1992,Nooijen1993}.  However, most of these methods do not
provide the forward and backward time-orderings needed to entirely subsume the
conventional RPA; two notable exceptions 
are the EOM-CC Green's function with single, double, and triple excitations, as
discussed recently in relation to the $GW$ approximation~\cite{Lange2018}, and
the Faddeev random-phase approximation~\cite{Barbieri2007,Degroote2011}.

\section*{Acknowledgments}

All calculations were performed with the PySCF software package~\cite{Sun2018},
using resources provided by the University of Chicago Research Computing Center.
This work was supported by the Air Force Office of Scientific Research under
award number FA9550-18-1-0058.
T.C.B.~is an Alfred P.~Sloan Research Fellow.
The Flatiron Institute is a division of the Simons Foundation.

\providecommand{\latin}[1]{#1}
\makeatletter
\providecommand{\doi}
  {\begingroup\let\do\@makeother\dospecials
  \catcode`\{=1 \catcode`\}=2 \doi@aux}
\providecommand{\doi@aux}[1]{\endgroup\texttt{#1}}
\makeatother
\providecommand*\mcitethebibliography{\thebibliography}
\csname @ifundefined\endcsname{endmcitethebibliography}
  {\let\endmcitethebibliography\endthebibliography}{}

\end{document}